# Neural network-based on-chip spectroscopy using a scalable plasmonic encoder

Calvin Brown[1,†], Artem Goncharov[1,†], Zachary Ballard[1,2,†], Mason Fordham[1], Ashley Clemens[3], Yunzhe Qiu[1], Yair Rivenson[1,2,4], and Aydogan Ozcan[1,2,4,5,*]

[1]*Department of Electrical and Computer Engineering, University of California, Los Angeles, California 90095, United States*

[2]*California NanoSystems Institute (CNSI), University of California, Los Angeles, California 90095, United States*

[3]*Department of Chemistry and Biochemistry, University of California, Los Angeles, California 90095, United States*

[4]*Department of Bioengineering, University of California, Los Angeles, California 90095, United States*

[5]*Department of Surgery, David Geffen School of Medicine, University of California, Los Angeles, California 90095, United States*

[†]These authors contributed equally to this work

*ozcan@ucla.edu

# Abstract

Conventional spectrometers are limited by trade-offs set by size, cost, signal-to-noise ratio (SNR), and spectral resolution. Here, we demonstrate a deep learning-based spectral reconstruction framework, using a compact and low-cost on-chip sensing scheme that is not constrained by the design trade-offs inherent to grating-based spectroscopy. The system employs a plasmonic spectral encoder chip containing 252 different tiles of nanohole arrays fabricated using a scalable and low-cost imprint lithography method, where each tile has a unique geometry and, thus, a unique optical transmission spectrum. The illumination spectrum of interest directly impinges upon the plasmonic encoder, and a CMOS image sensor captures the transmitted light, without any lenses, gratings, or other optical components in between, making the entire hardware highly compact, light-weight and field-portable. A trained neural network then reconstructs the unknown spectrum using the transmitted intensity information from the spectral encoder in a feed-forward and non-iterative manner. Benefiting from the parallelization of neural networks, the average inference time per spectrum is ~28 µs, which is orders of magnitude faster compared to other computational spectroscopy approaches. When blindly tested on unseen new spectra (N = 14,648) with varying complexity, our deep-learning based system identified 96.86% of the spectral peaks with an average peak localization error, bandwidth error, and height error of 0.19 nm, 0.18 nm, and 7.60%, respectively. This system is also highly tolerant to fabrication defects that may arise during the imprint lithography process, which further makes it ideal for applications that demand cost-effective, field-portable and sensitive high-resolution spectroscopy tools.

# Introduction

Spectral analysis is used in a wide array of applications in the fields of chemistry, physics, and biomedical sensing, among others. Optical spectra are conventionally recorded with spectrometers that separate light into its spectral components via a diffraction grating. The intensity of each component is recorded by a photodetector array, e.g., a complementary metal–oxide–semiconductor (CMOS) imager, to translate these intensities into the optical spectrum of the illumination beam (covering e.g., 400–750 nm). The Czerny-Turner configuration, for example, is one of the most commonly used methods for optical spectroscopy, employing two collimating mirrors to fold the optical path while partially compensating for optical aberrations[1]. Though elegant and robust, grating-based designs present two key performance trade-offs. Firstly, increasing the spectral resolution generally comes at the cost of decreasing the signal-to-noise ratio (SNR). For example, narrowing the entrance slit width, decreasing the period of the grating, or decreasing the pixel size of the sensor all improve spectral resolution at the expense of signal strength[2]. These methods also necessitate more expensive components and more precise instrument alignment. Such trade-offs can be prohibitive for low-light, low-cost, or field-based applications that still demand high spectral resolution. Secondly, increasing the spectral resolution may require a longer optical path between the grating and the photosensor array[3]. This is typically achieved with physically larger instruments (benchtop-sized), which are less suitable for mobile spectroscopy applications. In addition, a longer path length can degrade performance due to even minor ambient temperature fluctuations[2]. Therefore, traditional spectrometer designs present a compromise among resolution, cost, size, and SNR.

Computational sensing schemes have been proposed as a promising alternative to conventional grating-based spectrometers, presenting a variety of hardware and software solutions[4–11]. Instead of relying on diffraction gratings, some of these earlier systems work by encoding the incident spectra over a set of unique filter functions. The encoded information is then interpreted by a spectral reconstruction algorithm that employs precise a priori knowledge of the filter functions or leverages some calibration data to map the encoding operation to the target spectral measurement. Benefiting from the computational sensing paradigm, these emerging designs do not share the same size, throughput, and resolution trade-offs inherent to grating-based spectrometers. The quality of the spectral reconstruction is not explicitly linked to the optical path length or the spectral resolution of the detection scheme since the encoding operation does not divide the incident light into its narrowband spectral components, instead samples the input spectra with filters that can exhibit broadband transmission[5,12,13]. Performance of these computational schemes for spectroscopy therefore depends on the robustness and spectral diversity of the encoding operation as well as on the accuracy and speed of the employed algorithm to solve the underdetermined reconstruction problem[4,12,14,15].

A number of different hardware approaches have been proposed for the spectral encoding operation including variable filters in the form of liquid crystals and Fabry-Perot cavities as well as fixed filter configurations like ring resonators, Mach-Zehnder Interferometers (MZIs), photonic crystals and plasmonic filters[7–9,16–25]. Each encoding element, which may range from a narrowband spectral filter to a broadband filter function with multiple local extrema, samples the input spectrum $I(\lambda)$ using the filter functions of the spectral encoder. Reconstruction algorithms are therefore tasked to recover the incident spectrum from the raw data sampled by each encoder. The most common approach to algorithmic reconstruction is to use a priori information of the encoding operation and spectral sensitivity of the photodetectors to define a transformation, $T_i(\lambda)$, between the target spectrum, $I(\lambda)$, and raw measurements, $S_i$, i.e. $S_i = T_i(\lambda)I(\lambda)$ for each $i^{th}$ encoding operation. By expressing this transformation operation over all the encoding elements, a least-squares problem can be defined, and a solution for $I(\lambda)$ can be obtained by minimizing e.g., $\|S - TI\|_2^2$. Regularization terms based on the L1 norm (least absolute shrinkage operator, LASSO[26]) and the L2 norm (Tikhonov regularization[27]), among others, are also commonly used to solve this minimization problem, necessitating iterative reconstruction algorithms that overcome the limitations of the standard least-square solution to this underdetermined problem[5–7,9,10,15,16,18,19,24,28,29]. However, given this body of work, a data-driven non-iterative spectral reconstruction approach, without the need for a priori knowledge of the specific filter functions, has yet to be demonstrated.

Here we report a deep learning-based on-chip spectrometer (Fig. 1) that utilizes a flat spectral encoding chip (fabricated through a scalable and low-cost imprint lithography process) to filter the incident light using an array of 252 nanostructured plasmonic tiles, where each tile has a unique transmission spectrum. The transmitted light through all these tiles is acquired in parallel using a conventional CMOS image sensor that is axially positioned at ~3mm away from the plasmonic encoder, recording the free-space diffraction patterns of the plasmonic encoder without any lenses or optical components, using a compact and field-portable design (Fig. 1b). A trained neural network is used to reconstruct the unknown input spectra from the lensfree diffraction images in a feed-forward (i.e. *non-iterative*) manner without the need for a priori information on the encoding operation or the input illumination (Figure 1d,e). By leveraging batch computation, the network generates spectral reconstructions in ~28 µs per spectrum, which is orders-of-magnitude faster than other computational spectroscopy methods. When blindly tested on new input spectra of varying complexity (N = 14,648) captured *after* the training phase, the deep learning-based on-chip spectrometer correctly identified 96.86% of the spectral peaks with a peak localization error of 0.19 nm, a peak height error of 7.60%, and a peak bandwidth error of 0.18 nm. These performance metrics demonstrate significant improvements compared to earlier generations of computational spectrometers and were achieved despite visible fabrication defects in the plasmonic encoder chip, illustrating the robustness of our neural network-based spectral reconstruction method.

Taken together, the presented on-chip plasmonic spectroscopy design is highly cost-effective, compact, field-portable and requires no mechanical scanning components (Fig. 1). The methods and the device design that are at the heart of this computational on-chip spectrometer can find unique applications in various fields that demand compact and sensitive high-resolution spectroscopy tools.

## Results

**On-chip spectroscopy framework and experimental set-up**

Our deep learning-based spectral reconstruction framework (Figure 1a,b) uses a spectral encoding chip comprising 14 × 18 = 252 unique plasmonic 'tiles', where each tile covers a region of 100 × 100 µm, defined by a unique nanohole array structure (Figure 2). Importantly, the encoder chip is fabricated through a scalable imprint lithography process (Figure 2a) that can replicate nanostructures indefinitely from a silicon 'master' chip (see Methods). As a result, our encoding chip is low-cost and with the exception of a metal deposition step, can be fabricated without the need for clean room instrumentation or processes. Each plasmonic tile, *i*, serves as a unique spectral filter described by a transmission function, $T_i(\lambda)$, where the local maxima and their corresponding bandwidths result from the plasmonic modes supported by the dielectric and metal nanostructures (see Figure 2d,e).

For each illumination spectrum under test, the lensfree diffraction images of the input radiation were captured at multiple exposures to create a high dynamic range (HDR) image to limit pixel saturation effects. Lensfree images of the encoder chip corresponding to spectral peaks from 480–750 nm are shown in Supplementary Movie 1. Each of the 252 tiles is automatically segmented and further subdivided into a 9 × 9 grid of 81 sub-tiles. The average pixel intensities of all 252 × 81 = 20,412 sub-tiles serve as the input to a trained neural network, which rapidly reconstructs the unknown illumination spectrum without any iterations, in a single feed-forward manner (Figure 1c,d). The spectral reconstruction neural network comprises three fully-connected layers of 2,048 neurons each, and an output layer with 1,185 nodes, representing the spectral intensities over a target spectral range of 480–750 nm, with a spectral spacing of 0.229 nm. The network was trained using a mean squared error (MSE) loss function between the reconstructed spectra (network output) and the ground truth spectra, measured by a commercially available spectrometer (see Methods). To train the network, 50,352 random spectra were generated by a programmable supercontinuum laser, with an additional 8,824 spectra used for validation data (see Methods). The resulting neural network was blindly tested on 14,648 unseen new spectra generated by the same experimental set-up. Because the network requires no further training or iterations, it is able to reconstruct new, unseen spectra in ~28 µs on average in a single feedforward manner using a desktop computer (see the Methods section).

**Blindly tested spectral reconstructions**

Figure 3 illustrates the success of the trained reconstruction network to accurately recover unknown spectra using lensfree diffracted images that are acquired by our compact set-up (Fig. 1). The average MSE, peak localization error, peak intensity error, and bandwidth estimation error on the blindly tested spectra were 7.77e-5, 0.19 nm, 7.60%, and 0.18 nm, respectively. Overall, our experimental results reveal that 96.86% of the peaks in the ground truth spectra were correctly reconstructed by the network. Figures 3e-j further show reconstructed spectra and ground truth spectra for both lower complexity (one peak) and higher complexity (4–8 peaks) spectra for various performance percentiles. These percentiles refer to the MSE loss of the network output reconstruction, where $10^{th}$ percentile implies a relatively good fit (best 10% loss), $50^{th}$ percentile implies the median fit, and $90^{th}$ percentile implies a poor fit (worst 10%). Even for higher complexity spectra, the $90^{th}$ percentile network output results are rather accurate, closely matching the ground truth spectra acquired with a benchtop spectrometer. Additional examples of blind spectral reconstructions obtained at the network output are shown in Supplementary Figures 1–8 to demonstrate the repeatability and success of this blind spectral inference process.

We also evaluated the peak localization and bandwidth estimation error, on blindly tested new spectra, each with a 3 nm-bandwidth peak, ranging from 480–750 nm with a step size of 1 nm. For these 271 new spectra, all the peak localization errors were within ± 0.32 nm, and all the bandwidth estimation errors were within ± 0.178 nm, significantly surpassing the performance of earlier on-chip spectroscopy results.

**Spectral inference stability as a function of time**

Because the training spectra were captured *before* the blind testing, one would expect some level of performance degradation in spectral inference due to e.g., temperature fluctuations, illumination source stability/coupling changes, or mechanical vibrations, especially as the time period between the capture of the training and testing data increases. The performance stability of the inference of the trained network over the course of the blind testing data capture (~15 h of *continuous* operation) is evaluated in Figure 4. All performance metrics remained fairly stable, with no significant difference between their values at the start and end of the 15 h continuous testing period (Figure 4).

To further investigate the performance stability over time, an additional 21,296 new spectra were captured ~5.8 days *after* the last training/validation spectrum. Compared with the earlier blind inference results, the performance relatively degraded on these new spectra, as shown in Figure 5 (green curves). The average MSE, peak localization error, peak intensity error, and bandwidth estimation error on these later-acquired unknown spectra were 6.89e-4, 0.53 nm, 14.06%, and 0.29 nm, respectively, with 94.97% of spectral peaks correctly identified. As a means to re-calibrate the reconstruction network and overcome this relative performance degradation over time, we implemented a transfer learning approach, where the weights of the previously-trained neural network were adjusted through further training on a small fraction of the spectra captured at the start of the new measurement period (i.e., ~5.8 days *after* the last training phase). The performance metrics and spectral reconstructions after this transfer learning step are shown alongside those of the original network in Figure 5. All performance metrics are significantly improved after the transfer learning step: average MSE, peak localization error, peak intensity error, and bandwidth estimation error improve to 3.68e-4, 0.42 nm, 10.83%, and 0.23 nm, respectively, with 96.37% of the peaks correctly identified. Figure 5 further illustrates that, in addition to these considerable improvements in spectral inference metrics, background spectral noise and erroneous peaks are also suppressed well, after the transfer learning step.

It is important to emphasize that the amount of data and the computation time required for this transfer learning step are rather small; even using just 100 new spectra (requiring ~6 min to capture) and training the existing neural network for 100 epochs (requiring < 1 min on a desktop computer) shows marked improvements in the blind spectral reconstructions after 5.8 days (Supplementary Figure 12). Therefore, transfer learning can be an effective software-based calibration tool for our data-driven computational on-chip spectrometer, as demonstrated here.

**Speed of spectral reconstructions**

Unlike optimization-based approaches to spectral reconstruction[5,7,9,10,15,16,18,19,24,28,29], our neural network-based inference does not require iterative computation to predict each unknown spectrum. Once the network has been trained, it can perform reconstruction of unseen spectra rapidly and in parallel. The average prediction time per spectrum for different batch sizes are shown in Figure 6. All calculations were performed on a desktop computer (see Methods). For a batch size of 4096 spectra, the network is able to reconstruct an unknown spectrum in ~28 µs on average, providing orders of magnitude inference speed advantage compared to optimization-based iterative reconstruction methods. This parallel computation capability could be particularly beneficial for e.g., significantly increasing the speed of hyperspectral imaging systems, where a unique spectrum needs to be reconstructed for each hyperspectral pixel.

Compared to other state of the art neural networks used for image processing and enhancement tasks[30–35], the spectral reconstruction network employed in this work is compact and shallow, comprising three hidden layers and no convolutional layers. This enabled rapid prediction on an unseen spectrum in ~43 µs, without requiring an iterative minimization in the blind inference phase. To further increase the speed of prediction, we investigated subsampling each tile into a $7 \times 7$ sub-grid instead of $9 \times 9$ (Supplementary Figure 11). While the coarser subsampling ($7 \times 7$) causes a modest degradation in prediction performance, the network inference time further decreased to ~18 µs per spectrum. It is also important to note that the spectral reconstruction network yielded the best performance when trained with a dropout probability of 3%, much lower than the typical values (10–50%) used in many common neural networks employed in other applications[30–32,36]. Because the neural network does not use convolutional layers and is relatively shallow and wide, even a small dropout probability gave us a strong regularization effect.

**Network generalization to new spectral distributions not included in the training**

In addition to investigating the spectral reconstruction network's ability to generalize over time, we also tested its ability to generalize to new spectral distributions that were not included in the training phase. While the network was trained on random spectral distributions containing 1–8 peaks, we synthesized more complex spectra that had more number of peaks, not represented within the training phase. The network was tested on these synthesized spectra without any retraining or transfer learning, and some examples of the synthesized spectra and the corresponding spectral reconstructions are shown in Fig. 1e and Supplementary Figure 9. Despite using spectral distributions that were never represented during the training phase, the reconstructions still identify 90.54% of all peaks and do not suffer from any obvious artifacts or noise. These results demonstrate the presented framework's ability to generalize to more complex spectral distributions that were not included in the training data.

# Discussion

The performance of the presented spectral reconstruction framework demonstrates its potential to enable new modalities of spectroscopy and hyperspectral imaging. When our deep learning-based spectrometer was blindly tested on unseen spectra, it correctly identified 96.86% of the spectral peaks, with a peak localization error of 0.19 nm, a peak height error of 7.60%, and a peak bandwidth error of 0.18 nm. The presented spectral reconstructions (Figure 3, Supplementary Figures 1–8) match the ground truth spectra quite well, both quantitatively and qualitatively, with the peak localization error and the peak bandwidth error on the order of the resolution (0.229 nm) of the spectrometer used to capture the ground truth training data. Additionally, there is no observed bias in these performance metrics over the operational range (480 nm – 750 nm) which can be partly attributed to (1) the lack of wavelength dependency in the MSE loss function, and (2) the spectral diversity of the filter functions, $T_i(\lambda)$, of the plasmonic encoder chip (see Figure 2e). It is also important to emphasize that as a unique aspect of this work, successful reconstruction performance was demonstrated over a large blind testing dataset (14,648 unique spectra) containing varying degrees of complexity, including non-sparse examples with overlapping spectral features (e.g., Figure 3h,j). When compared to a linear regression model with L2-norm regularization trained on the same data set, the trained neural network achieved nearly a 5-fold reduction in average MSE (from 3.85e-4 to 7.77e-5). The predicted spectra resulting from the regularized linear model contain significant noise artefacts (see Supplementary Figure 10) not seen in our deep-learning based reconstructions, suggesting that the hidden layers and the inherent nonlinearities within the neural network play an important role in de-noising the spectral features identified through linear operations, yielding robust performance over complex and non-sparse inputs.

Avoiding overfitting of the spectral reconstruction algorithm is critical for any computational spectrometer that uses a data-driven approach. This is especially important for complex models such as neural networks that contain a large number of trainable parameters and non-linear activation functions. Reconstruction algorithms that exhibit overfitting to a particular training set, can fail to appropriately interpret minute changes in experimental system alignment, temperature, vibrations, or other unforeseen noise sources, potentially leading to significant changes in the output and overall reconstruction performance. The presented system showed very good stability over the course of ~15h *continuous* experimentation during which the 14,648 blind testing spectra were captured (Figure 4); however its blind inference performance relatively degraded after ~5.8 days, likely due to uncontrolled factors, e.g., temperature, vibrations. These spectral reconstructions after ~5.8 days were improved considerably by transfer learning on newly-captured data, which amounts to a simple calibration step, similar to what is typically used for some other measurement instruments.

Another important aspect of this spectral reconstruction framework is the use of a spectral encoder chip, fabricated through scalable imprint lithography. The fabrication of the encoding chip does not require cleanroom-based lithography or other processes that require an advanced fabrication infrastructure, except the metal deposition step, which is relatively much simpler and cheaper. While this low-cost and rapid imprint lithography process can introduce point imperfections in the encoder chip, as evident in Supplementary Movie 1, the data-driven spectral reconstruction network demonstrates robustness to these defects. Due to the minimal cost and scalability of the imprint lithography, large area encoders for hyperspectral imaging could be fabricated, in which an ensemble of optimal filter functions could be grouped into a single hyperspectral pixel that is tiled across the encoder chip. While the present encoder contains 252 spectrally overlapping broadband filters, further optimization in its design can be achieved: application-specific feature reduction approaches can be used to select, in a data-driven manner, a statistically optimal sub-set of tiles[13,37]. In Supplementary Figure 13, we explored network performance when using random subsets of the 252 tiles, demonstrating that blind spectral reconstructions using just 49 plasmonic tiles are still quite competitive. This trade-off between reconstruction performance and the number of encoder elements would be critical for designing future computational spectrometers.

Compared to traditional grating-based spectrometer designs, the presented spectral reconstruction framework offers several unique features. First, the compact nature of the on-chip spectroscopy system (Figs. 1a,b) could enable inexpensive, lightweight designs with large fields-of-view for, e.g. remote, airborne, or even disposable sensing needs in field settings. Because the encoder chip can be manufactured at low cost over large areas with the imprinting process, an array of spectrometers or a hyperspectral imaging grid could be fabricated without the need for cleanroom-based lithography tools. Since the presented device bins the neighbouring pixels, spectrometers using large-pixel size sensors or, conversely, spectrometers with even a smaller footprint (via less pixel binning) could be designed as well. Second, the traditional trade-off between the spectral resolution and SNR that is common in grating-based spectrometers is now pushed to a different optimum point: the resolution of our spectral reconstruction network is primarily limited by the spectral resolution of the instrument used for ground truth measurements of the training data, and the individual filters of the encoder chip do not need to be narrowband to match the ground truth resolution as demonstrated in this work.

The data-driven approach utilized in this work also offers key advantages when compared to common spectral reconstruction algorithms based on e.g., least-squares minimization employed in other computational spectrometer systems. Although the training process requires a large amount of measurements to be obtained, this is a one-time effort, and it yields a forward model that can blindly recover unknown spectra from raw sensing signals in ~28 μs, orders of magnitude shorter than the time required to solve iterative minimization problems, employed in earlier spectral reconstruction methods. Spectral reconstruction timing can be important for various applications such as hyperspectral imaging, that may demand a spectral recovery across a large sequence of images each with a large number of individual pixels. Additionally, some of the iterative reconstruction algorithms used earlier employ a 'smoothness' constraint in their optimization process, based on the second derivative of the target spectra[5]. Although this may improve some spectral reconstructions, the selection of a singular weighting parameter on this constraint introduces a trade-off in performance between narrow-band and broad-band spectral reconstructions. Lastly, instead of using training data, these iterative reconstruction methods rely on precise measurements of the encoding operation and the spectral response of the underlying photosensors, which are both used as a priori information. This presents a separate array of challenges, because the reconstruction performance relies on how precisely one can characterize the underlying hardware. All of these challenges are considerably mitigated or eliminated using the presented deep learning-based spectroscopy approach, which also lends itself to a highly compact, field-portable, sensitive and high-resolution spectrometer design that can be used in various targeted applications, in both sensing and imaging.

## Methods

**Plasmonic encoder**

The plasmonic encoder measures 4.8 × 3.6 mm and consists of 252 (14 × 18) tiles, with each tile covering 100 × 100 μm. Each one of these tiles consists of a nanohole array with a unique combination of periodicity (square or hexagonal), period (280–650 nm), and aspect ratio (period divided by hole diameter, spanning 1.5–3.0) (Figure 2b,c). As a result, the 252 plasmonic tiles support distinctive plasmon resonances in the visible range of optical spectrum, manifesting as unique filter functions for the incident light.

The embedded nanostructures in the encoder are molded from a silicon 'master' chip that contains the desired nanohole array designs. The silicon master was fabricated using Electron-beam lithography (Raith EBPG5000 ES) with a ZEP520A resist (Supplementary Figure 14). After the resist was exposed and developed, a Chlorine etcher (ULVAC NE 500 with 5 sccm Ar, 20 sccm $Cl_2$) was used to create the nanohole arrays in the silicon. After the production of the master, plasmonic encoder chips were then fabricated using an imprint molding process described earlier[38]. The final encoder chip is comprised of a UV-curable polymer NoA-81 (Norland Products, Inc.) backed by a standard microscope slide, with 50 nm of gold and 5 nm Titanium adhesion layer deposited via Electron Beam Evaporation (CHA Solution).

**Experimental procedures**

Optical spectra were generated by a programmable supercontinuum laser (~3 nm bandwidth) with up to eight independent emission channels (Fianium, United Kingdom). Random spectra were created by turning on a random number of channels between 1 and 8. For each channel, the center wavelength was set randomly between 480 and 750 nm, and the power was set randomly between 0.1 and 0.7 (a.u.). All experiments were performed with random spectra. The output from the laser was coupled to a 50/50 2 × 1 fiber splitter (OZ Optics), with one arm coupled to the input aperture of our device and the other arm coupled to a conventional spectrometer (Ocean Optics HR+) to capture a ground truth spectrum for each measurement. For each spectrum, images were captured by the CMOS image sensor (daA1280-54um by Basler AG, Germany) at ten different exposure times (increasing in length by a factor of two each time) and the resulting images were combined into a single HDR image. Each spectrum was captured by the ground truth spectrometer five times and the resulting spectra were averaged to minimize the effects of noise. Spectra that were over- or under-saturated (in either the ground truth spectrum or the captured HDR image) due to randomness of peak location and power were removed from the dataset.

Training and validation spectra were captured over the course of ~3.75 days. The training dataset consisted of 50,352 spectra, while the validation dataset consisted of 8,824 spectra. Data for blind testing were captured immediately afterward, consisting of 14,648 spectra captured over 15 h of continuous operation of the system. Additionally, another blind testing dataset was captured starting ~5.8 days after the last training/validation spectra were captured.

**Spectral reconstruction network**

All images were registered to an orthogonal grid to account for gradual drift of the encoder chip relative to the CMOS active area. The $252 \times 9 \times 9 = 20,412$ sub-tile intensities were combined into a vector to serve as the input to the neural network. The spectral reconstruction network comprises three fully-connected layers of 2,048 neurons each, and an output layer with 1,185 nodes (shown in Supplementary Figure 15). Batch normalization and dropout layers (with a dropout probability of 3%) were used after each fully-connected layer to prevent overfitting to the training spectra. The network was trained using the Adam optimizer[39] with a learning rate of 1e-5. All spectral reconstructions were performed on a desktop computer with a Titan RTX graphics processing card (NVIDIA).

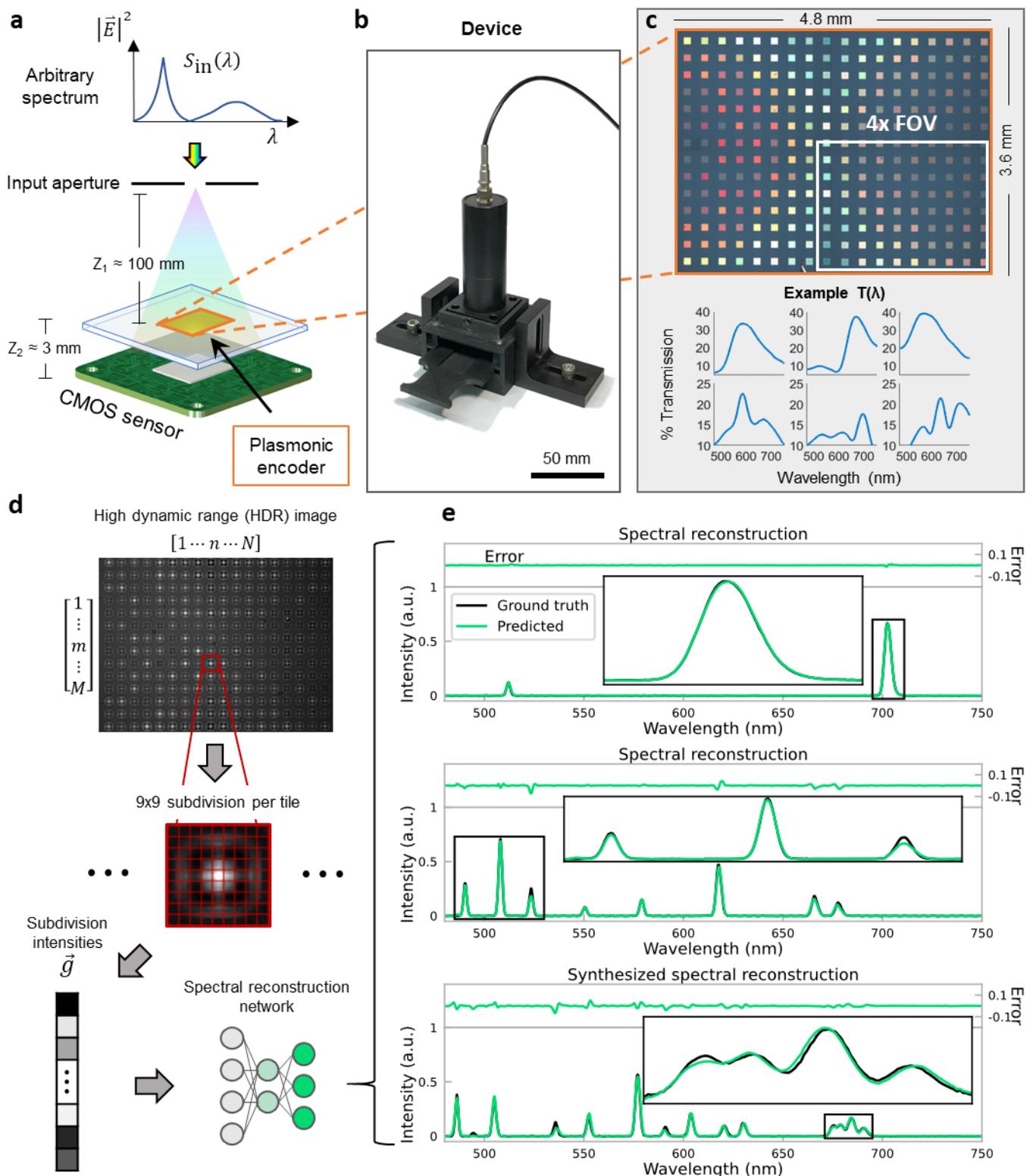

**Figure 1. Neural network-based on-chip spectroscopy. a** Schematic of the optical setup. Plasmonic encoder is located between a CMOS image sensor and the input aperture. **b** Photo of optical setup. **c** Brightfield microscope image of the plasmonic encoder chip showing example transmission spectra T(λ) below. **d** Workflow of spectral reconstructions. Regions of the HDR image corresponding to each tile are used as inputs to the spectral reconstruction neural network. **e** Spectra reconstructed during blind testing. Error is shown above each plot on the same y-scale. The network was trained only on spectra with up to 8 peaks, yet it successfully reconstructs a spectrum with 14 peaks.

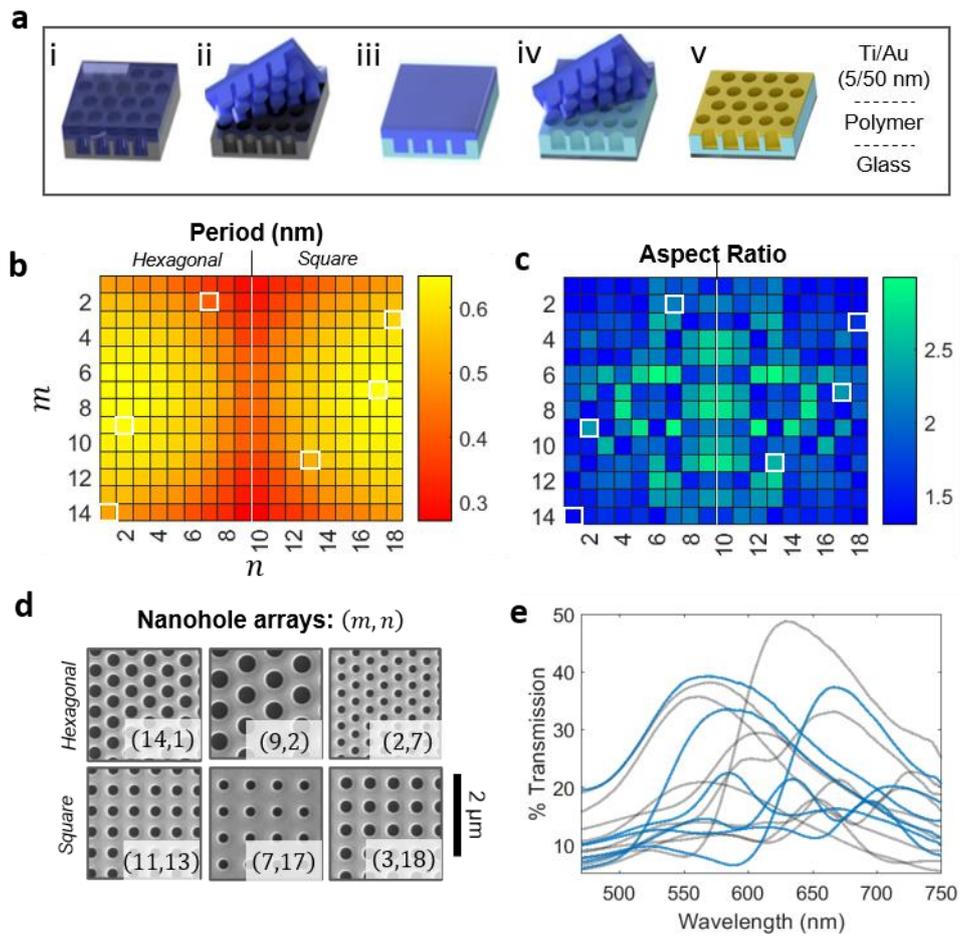

**Figure 2. Spectral encoder chip.** **a** Soft lithography process for molding low-cost replica of nano structures from a silicon master. Steps *i-ii* show the initial molding process with the silicon master and the initial UV curable polymer. Steps *iii-iv* show the secondary molding process followed by the metal deposition in ***v***. The **b** Period and **c** Aspect Ratio of the nanohole array for each tile in the encoder are shown using a heatmap. **d** SEM images of example plasmonic nanohole arrays, corresponding to the outlined white boxes in b and c and example spectra in Figure 1c. **e** Example transmission spectra, where the blue lines correspond to the plasmonic tiles shown in the SEM images and in Fig 1c. Other example transmission spectra are shown in grey.

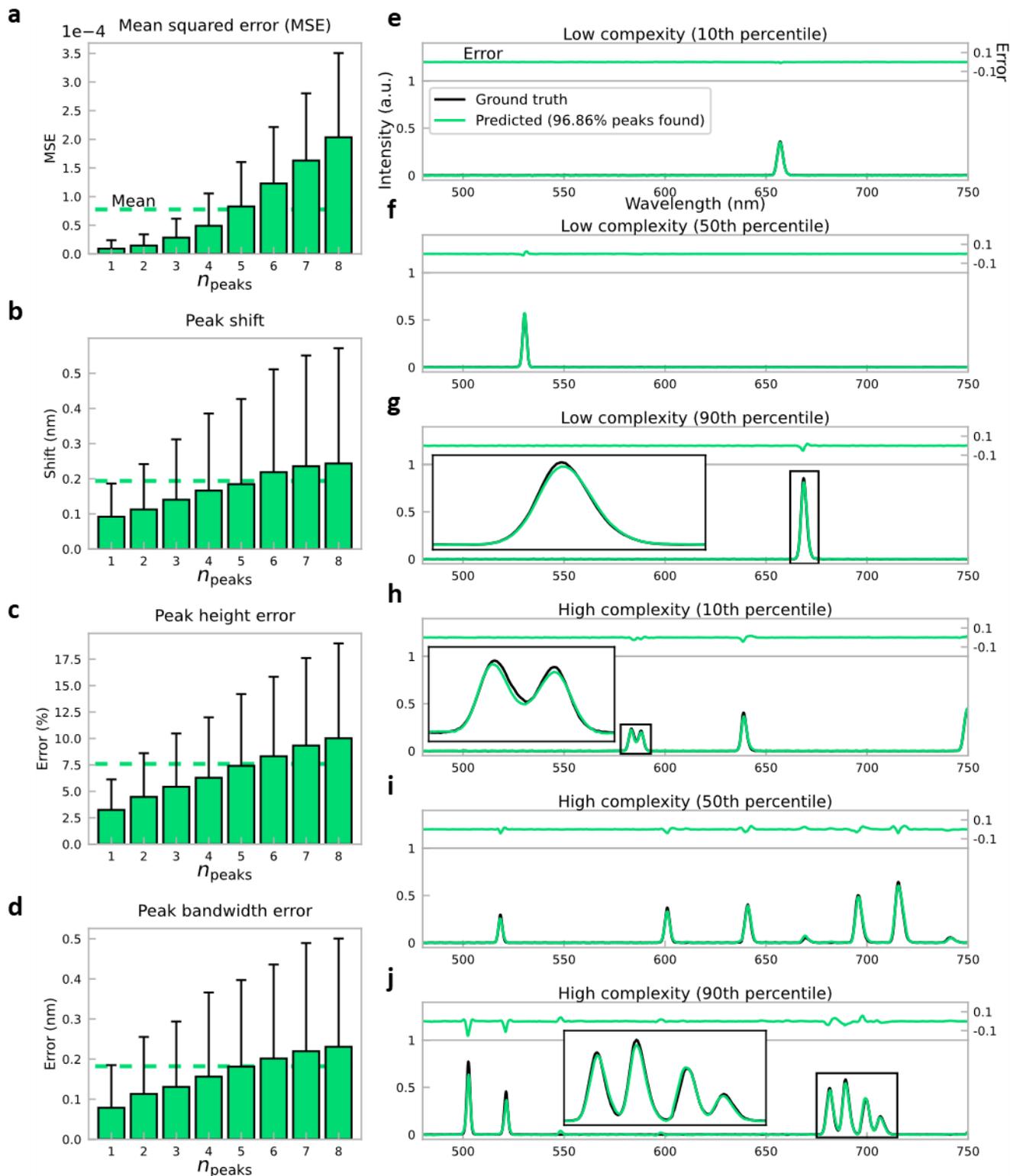

**Figure 3. Blind testing performance. a** Average MSE, **b** peak shift/localization error, **c** peak height error, and **d** peak bandwidth error for spectra containing 1-8 peaks. Average over all spectra is shown as a horizontal line in each plot. Reconstructions for lower complexity (1 peak) spectra in the **e** 10$^{th}$, **f** 50$^{th}$, and **g** 90$^{th}$ percentile of MSE. Reconstructions for higher complexity (4-8 peaks) spectra in the **h** 10$^{th}$, **i** 50$^{th}$, and **j** 90$^{th}$ percentile of MSE. 10$^{th}$, 50$^{th}$, and 90$^{th}$ percentiles correspond to best 10%, median, and worst 10% fits, respectively. Error is shown above each plot on the same y-scale.

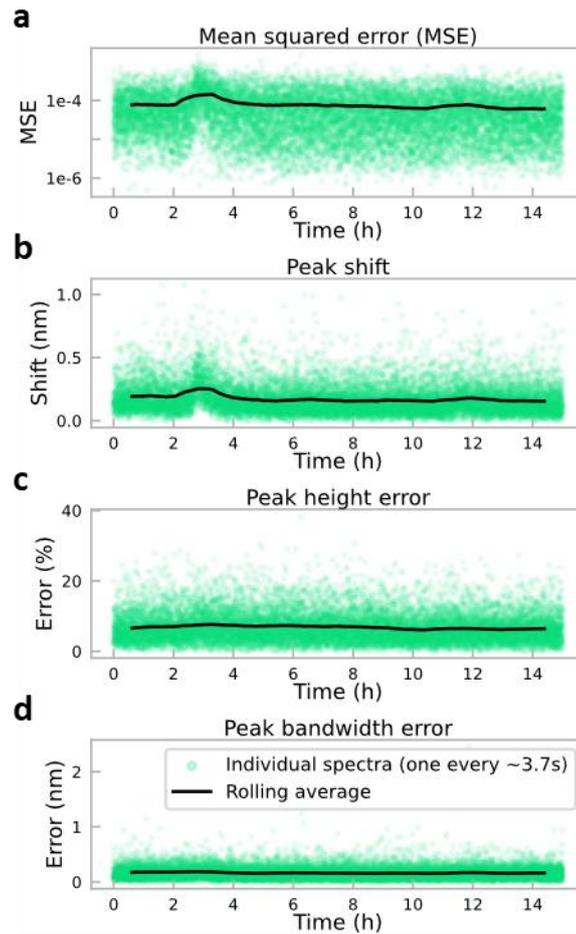

**Figure 4. Stability of inference performance over time. a** Average MSE, **b** peak shift/localization error, **c** peak height error, and **d** peak bandwidth error for spectra over the course of blind testing data capture. 14,648 blind testing spectra are represented in the plots, captured over ~15 h of continuous operation of the system.

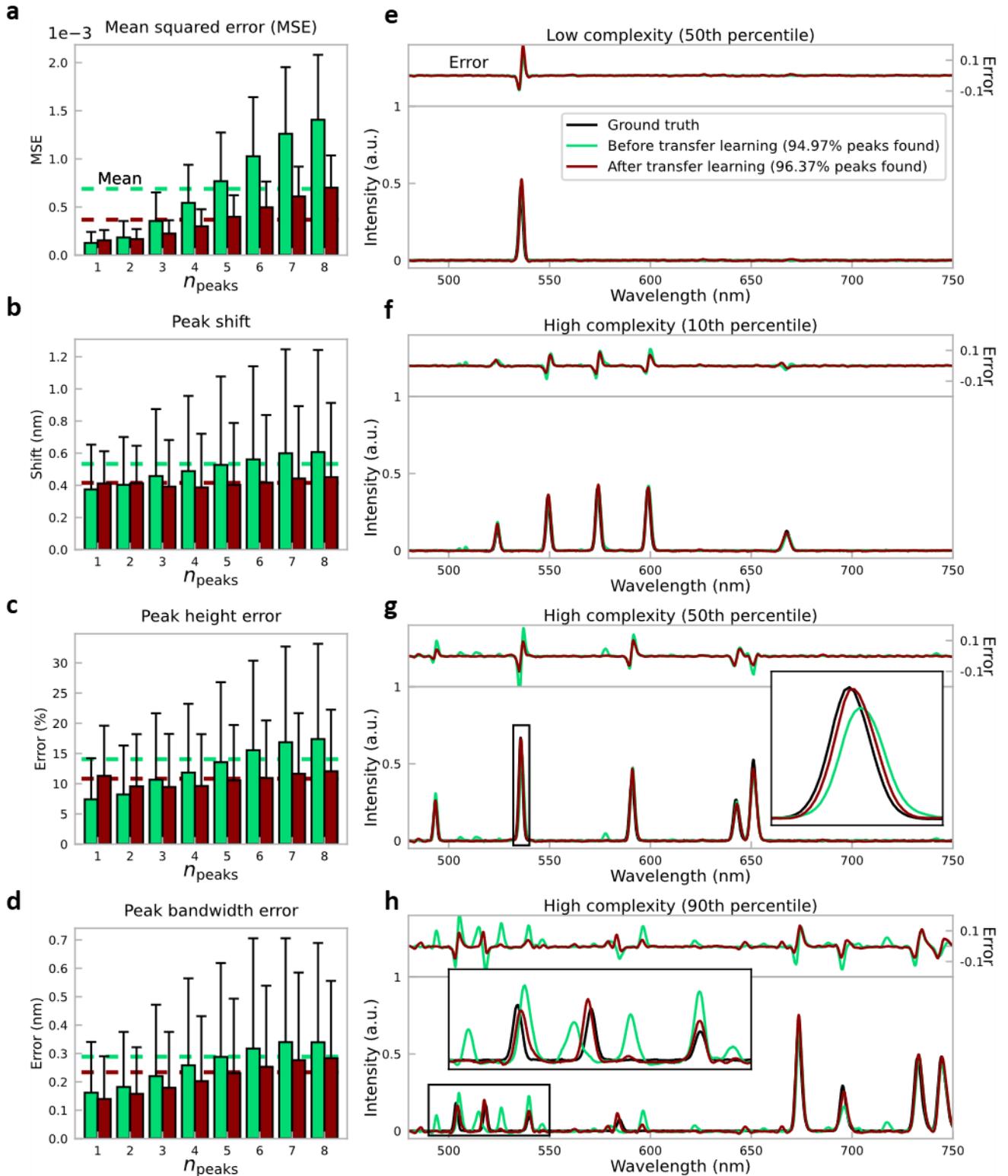

**Figure 5. Performance on blind testing spectra captured ~5.8 days after training. a** Average MSE, **b** peak shift/localization error, **c** peak height error, and **d** peak bandwidth error for spectra containing 1-8 peaks. Average over all spectra is shown as a horizontal line in each plot. Transfer learning considerably improves the inference performance, for all metrics. Reconstruction for lower complexity (1 peak) spectrum in the **e** 50th percentile of MSE. Reconstructions for higher complexity (4-8 peaks) spectra in the **f** 10th, **g** 50th, and **h** 90th percentile of MSE. Error is shown above each plot on the same y-scale.

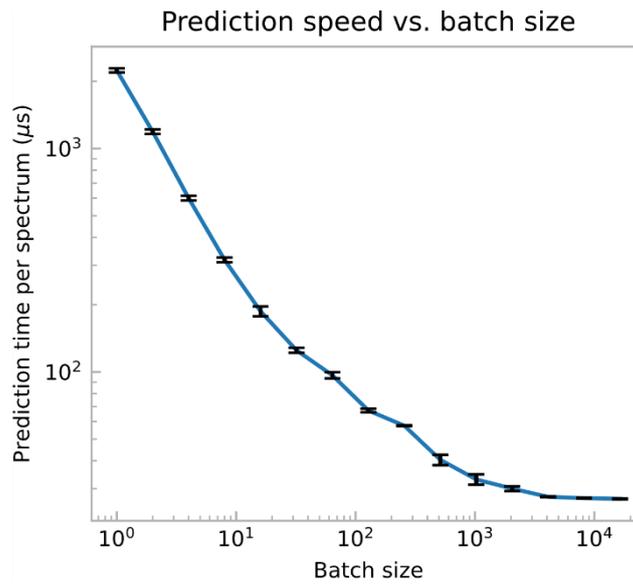

**Figure 6. Prediction speed vs. batch size for blind testing spectra.** Inference time per spectrum considerably decreases with increasing batch size due to the highly parallelizable nature of the neural network computation. An average inference time of ~28 μs is obtained for a batch size of ≥4096. Error bars were generated from 7 repeated trials. All predictions were performed on a desktop computer (see Methods).